\documentclass[aps,prb,twocolumn,floatfix,superscriptaddress,showpacs]{revtex4}

\usepackage{graphicx}
\usepackage{bm}

\begin{document}

\title{Theoretical optical and x-ray spectra of
liquid and solid H$_2$O}

\author{J. Vinson}
\affiliation{Dept.\ of Physics, Univ.\ of Washington Seattle, WA 98195}

\author{J. J. Kas}
\affiliation{Dept.\ of Physics, Univ.\ of Washington Seattle, WA 98195}

\author{F. D. Vila}
\affiliation{Dept.\ of Physics, Univ.\ of Washington Seattle, WA 98195}

\author{J. J. Rehr}
\affiliation{Dept.\ of Physics, Univ.\ of Washington Seattle, WA 98195}

\author{E. L. Shirley }
\affiliation{National Institute of Standards and Technology, Gaithersburg,
MD 20899}

\date{\today}

\begin{abstract}
Theoretical optical and x-ray spectra of model structures of water and ice
are calculated using a many-body perturbation theory, Bethe-Salpeter equation
(BSE) %, 
approach implemented in the valence- and
core-excitation codes AI2NBSE and OCEAN. These codes use 
{\it ab initio}
density-functional theory wave functions from a plane-wave, pseudopotential
code, quasi-particle self-energy corrections, and a BSE treatment of
particle-hole interactions.
The approach improves upon independent-particle methods through the inclusion
of a complex, energy-dependent self-energy and screened particle-hole
interactions to account for inelastic losses
and excitonic effects.  These many-body effects are found to be
crucial for quantitative calculations of ice and water spectra.

\end{abstract}

\maketitle

\section{Introduction}

 Recently there has been considerable interest and controversy surrounding the
connection between the local structure of water and ice and their observed
optical and x-ray spectra.\cite{Soper2010,Ph.Wernet05142004,PhysRevLett.96.215502,chen-2009,Car2004}  
Part of the difficulty lies in modeling the
structures of these complex, finite-temperature systems.  
Conventional super-cell
methods require large unit cells to treat proton-disorder in ice
and a configurational average adequate to represent a
statistical ensemble of water structures. 
Structural probes such as x-ray and neutron scattering do not provide
an unambiguous interpretation of local geometry.\cite{Wikfeldt2009}  Another part
of the difficulty lies in theoretical modeling.  Despite many attempts,
quantitative theoretical calculations of the optical and x-ray spectra of
these systems have proved to be notoriously difficult due to strong
non-local, self-energy, and excitonic effects. Thus the various 
theoretical methods that have been employed to date exhibit considerable
variation in their results, undermining a definitive interpretation.
\cite{Ph.Wernet05142004,PhysRevLett.96.215502,Car2004,PhysRevLett.100.107401,chen-2009,Leetmaa2010}

In an effort to address these issues, we present calculations of the
valence and core excitation spectra of well-characterized model ice and water
systems based on a recently developed approach utilizing
 the Bethe-Salpeter equation (BSE) and Hedin's GW approximation for the
quasi-particle self-energy (the acronym  GW refers to the product
of the one-electron Green's function $G$ and the screened Coulomb
interaction $W$),\cite{Onida02} with a pseudopotential plane-wave
basis.  This GW/BSE approach has been implemented in the AI2NBSE
and OCEAN packages for valence and core-level spectra
respectively.\cite{lawler:205108,ocean}
The method is advantageous compared to independent-electron
approximations,\cite{PhysRevLett.96.215502,Ph.Wernet05142004,Fister2009}
in that it provides a first-principles method for the inclusion of 
both quasi-particle and excitonic effects. These many-body
effects modify peak positions, widths,  and strengths, and hence they
are crucial for a quantitative treatment and interpretation of the spectra.
GW/BSE calculations have been carried out previously for both the
valence\cite{PhysRevLett.94.037404,PhysRevLett.97.137402}
and core\cite{chen-2009} spectra of water, with various
approximations for the screened particle-hole interaction. A key difference
is that our implementation uses a complex, energy-dependent GW
self-energy, in addition to the screened particle-hole interactions.
 
The remainder of this paper is as follows: The theoretical methods
are summarized in Sec.\ II.\ 
Results for two forms of ice are presented in Sec.\ III,
and for a model water system in Sec.\ IV.
Finally, Sec.\ V contains a summary and suggestions for further work.

\section{Theoretical Methods}

\subsection{GW/BSE Approach}

The theory behind the GW/BSE approach implemented in AI2NBSE and OCEAN for
valence and core excitations is described in detail
in Refs.\ \onlinecite{lawler:205108} and \onlinecite{ocean} respectively,
so only a short summary is included here.  The BSE represents the 
equation of motion of a particle-hole state, here an electron photo-excited
into the conduction bands from either the occupied valence bands (optical)
or a deep Oxygen 1s orbital (x-ray).
Briefly, the open-source plane-wave,
pseudopotential code ABINIT\cite{Gonze2009,Gonze2005} is used to calculate
%%%%%% NIST reviewer worried about possible disclaimer wrt ABINIT
both occupied and unoccupied Kohn-Sham density functional theory (DFT) states
of the ground-state
Hamiltonian, which serve as a basis for the BSE. This Kohn-Sham calculation
uses the Ceperley-Alder, Perdew-Wang local density
approximation for the exchange-correlation potential.\cite{LDA}
DFT orbitals only strictly describe a non-interacting system, but they can 
be good approximations to quasi-particle wavefunctions and 
perturbativley corrected, e.g., via Hedin's GW approximation.
In this work, first-order quasi-particle energy corrections are added through
the GW many-pole self-energy (MPSE) approximation of Kas et al.\cite{kas:195116}
In contrast to previous GW/BSE approaches for these systems,
\cite{PhysRevLett.94.037404,PhysRevLett.97.137402}
our MPSE is complex and energy-dependent, and thus accounts for both the
energy shifts and inelastic losses which stretch and damp the spectral features.  
This MPSE model is based on a many-pole fit to AI2NBSE calculations
of the valence loss spectrum $-{\rm Im}\, \epsilon^{-1}(\omega)$.

The optical and x-ray spectra, the latter including both x-ray absorption
(XAS) and non-resonant inelastic x-ray scattering (NRIXS), which is also
referred to as x-ray Raman spectra (XRS), are then calculated using the
National Institute of Standards and Technology (NIST) BSE solver in
AI2NBSE \cite{lawler:205108} and OCEAN \cite{ocean} respectively.
The core spectra in OCEAN use atomic core-level states and projector-augmented-wave
 (PAW) transition matrix elements.\cite{Blochl94}
 The BSE kernel includes both a screened direct attraction between the electron and hole
 and an unscreened repulsive exchange term.
The treatment of the particle-hole interactions is an important consideration
due to the weakly screened excitonic effects 
in water and ice.  In both our valence and core codes the screened Coulomb interaction $W$ is 
approximated as statically screened. A Hybertsen-Levine-Louie 
dielectric function is implemented in AI2NBSE, 
while OCEAN uses the random-phase approximation at short-range,
switching to a model dielectric function at
long range.\cite{Shirley2006,Soininen2001}  
In contrast, previous
approaches have used a variety of approximations ranging from 
a self-consistently screened DFT
core-hole \cite{PhysRevLett.96.215502,chen-2009}
to half-core-hole transition-state potentials.\cite{Ph.Wernet05142004}

\subsection{ Structures }
\begin{figure}
\begin{center}
\includegraphics[scale=0.3,angle=270]{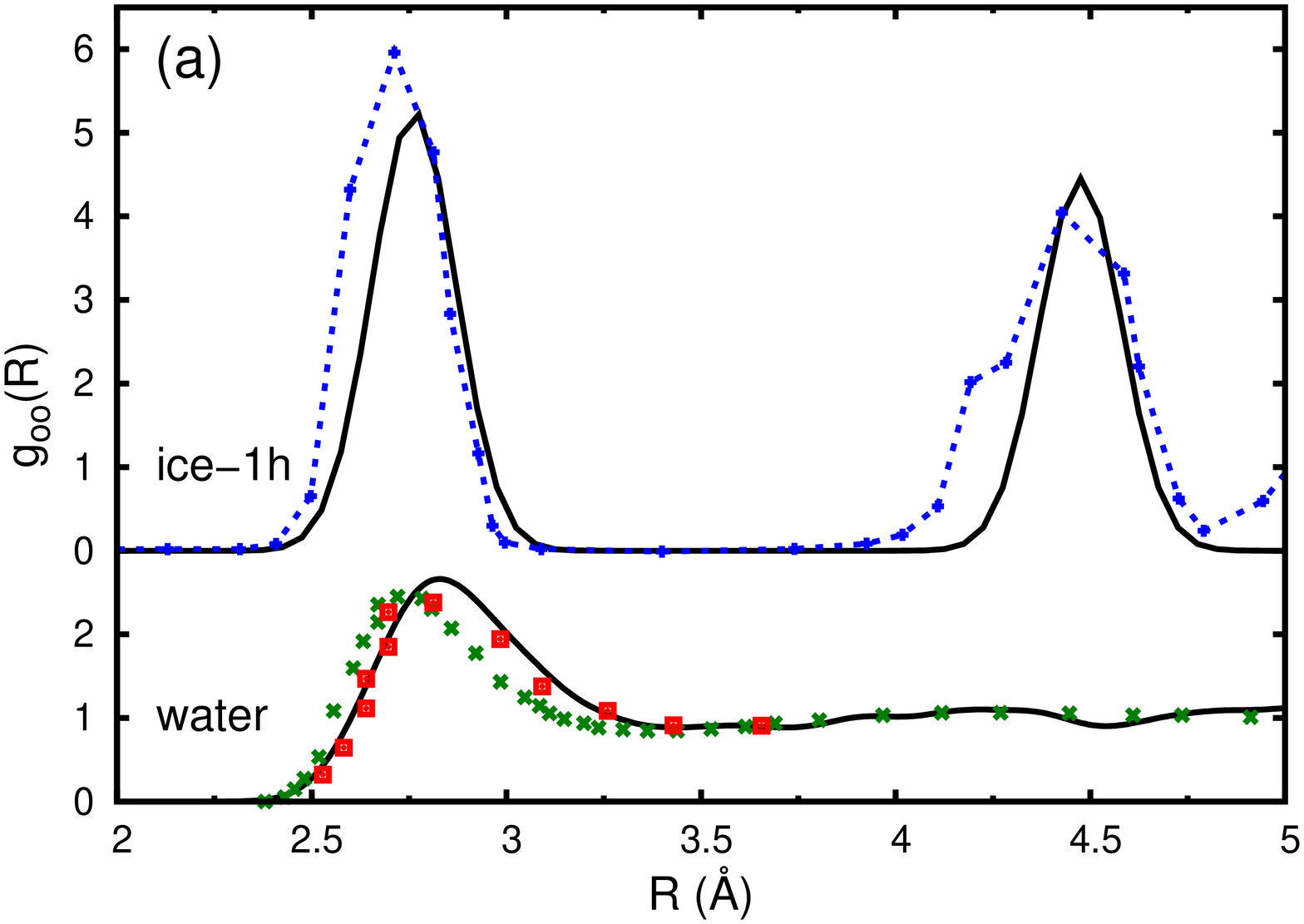}
\end{center}
\begin{center}
\includegraphics[scale=0.3,angle=270]{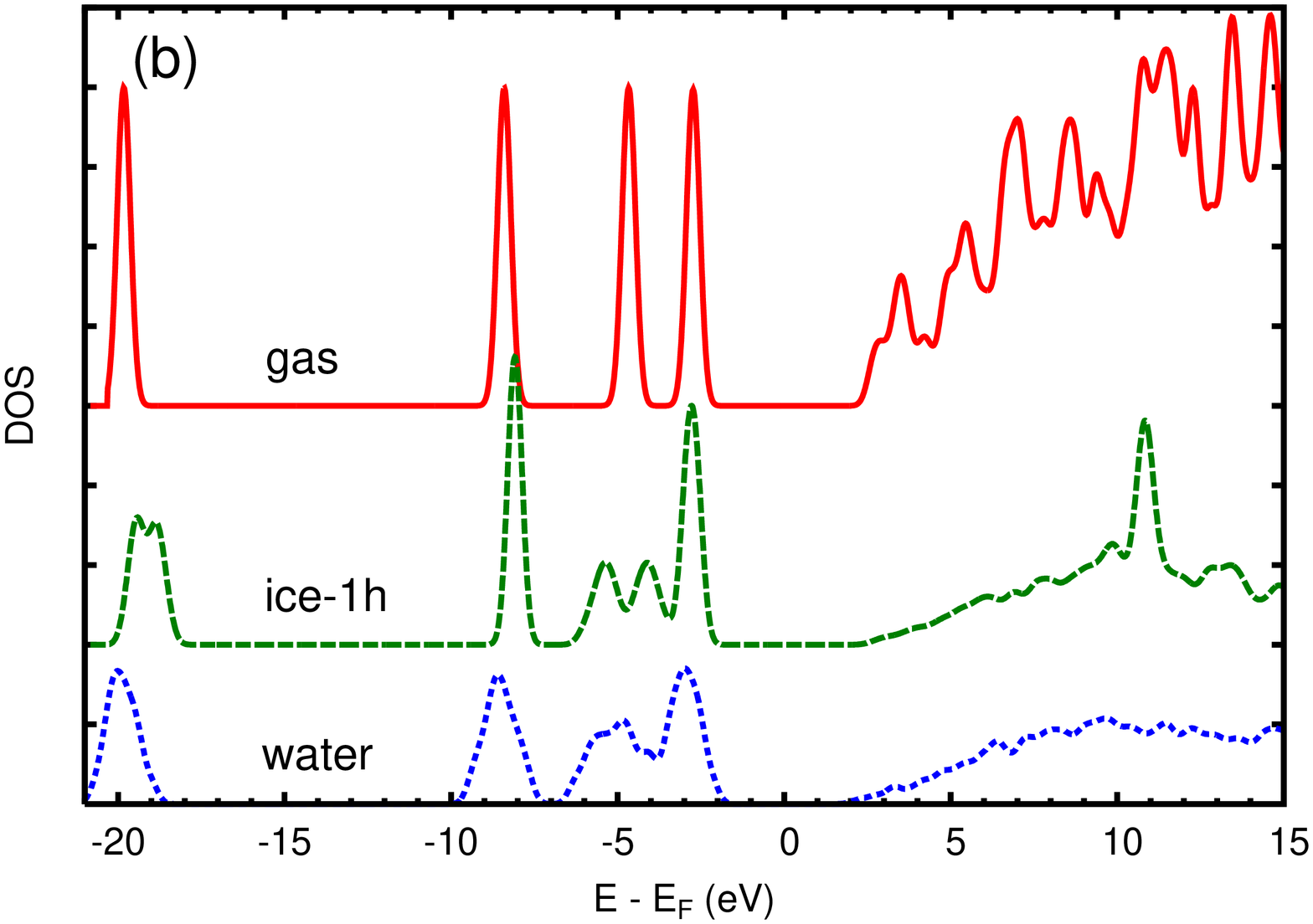}
\end{center}
\caption{(Color online) 
a) The oxygen-oxygen pair distribution function, $g_{OO}(R)$,
for our model structure
unit cells (black, solid line) compared with experimental results for
ice\cite{Stuart1981} (blue plusses) and two 
different water results (red boxes\cite{Soper1997} and green crosses \cite{Soper2007});
b) comparison of the total electronic density of states 
for ice-Ih, liquid water and gas-phase models, calculated using
ABINIT.  The broadening and vertical scaling
are the same for all three structures, and energies are relative to the Fermi
level. The features for the liquid cell are naturally broadened by
structural disorder.}
\label{DoS}
\end{figure}

The pair distribution function (PDF) $g(R)$, and total electronic
density of states
(DOS) provide useful measures of the ground-state structural and electronic
properties of our model ice and water systems (Fig.\ \ref{DoS}).
In order to interpret x-ray and optical spectra, it is important that the model structures used
in the calculations match observed structural properties, e.g.,
the PDF from neutron or x-ray diffraction.
For our model ice-Ih cell $g_{OO}(R)$ is in reasonable
agreement with experiment. However, the first shell peak is shifted to
a slightly larger mean radius.
To be consistent with experiment a root-mean-square
disorder of about 0.25~\AA\ is needed, which is simulated in
Fig.\ 1a by convolving the model PDF with a Gaussian.
For our 17-molecule water cells (see Sec.~III) fair agreement with
experiment is observed,
but with a slightly elongated first-shell O-O distance. The position and 
height of the first peak in $g_{OO}(R)$ are sensitive to both the 
momentum range in the diffraction measurements and assumptions about 
the structure and core potentials. Discussions of these limitations can be found elsewhere, e.g., Refs.\ 
\onlinecite{Wikfeldt2009} and \onlinecite{Soper2007}.

The total electronic density of states (DOS) provides a useful picture of the
ground-state electronic structure, and in particular, the occupied and
unoccupied
energy levels in these materials.  A comparison of the DOS for liquid,
solid, and gas-phase H$_2$O as calculated with ABINIT is shown in
Fig.\ \ref{DoS}b. The DOS of both liquid water and ice-Ih are roughly
similar, apart from broadening and the sharp peak at about +10 eV in
ice. The similarity among the occupied DOS suggests that the core states
are well localized and hence essentially molecular in character. 
Likewise, the similarity in the unoccupied DOS between liquid and
solid H$_2$O suggests that short-range order dominates the spectra. 
The additional broadening in the liquid
reflects the larger configurational disorder in amorphous structures.
In contrast, the unoccupied DOS of the gas phase shows considerably
more structure than either the liquid or solid phases. 
These results for the DOS are similar to previous work, e.g.,
Chen et al.\cite{chen-2009}

\subsection {GW Self-energy}

\begin{figure}
\includegraphics[scale=0.3,angle=270]{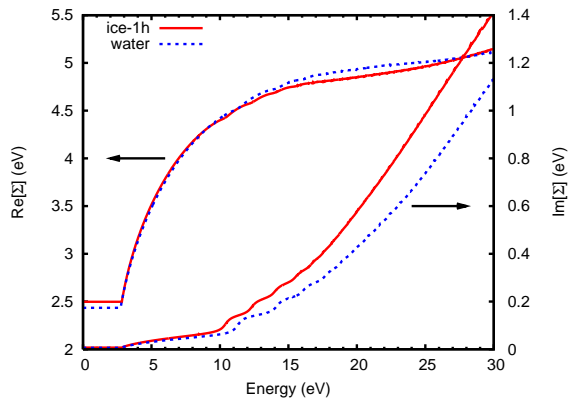}
\caption{(Color online)  Comparison of GW self-energies for solid and liquid
water as calculated with our MPSE model.\cite{kas:195116}
Note that these are quite similar, indicative of similar broadening and
self-energy damping effects in water and ice. The quasi-particle self-energy
broadening may be compared to other sources of broadening in the spectra: 
experimental resolutions ($\approx~0.5$~eV), the O 1s core-hole lifetime
($0.16$~eV), and vibrational broadening which is yet smaller and not included.  }
\label{SE_comp}
\end{figure}

Quasi-particle self-energy effects are of crucial importance in broad-spectrum
calculations of optical and x-ray spectra, because independent-particle
calculations systematically underestimate conduction band widths. 
A comparison of our GW many-pole self-energies for water and ice
is shown in
Fig.\ \ref{SE_comp}. Note that the imaginary parts are quite similar
and grow significantly above about 15~eV, while the real parts differ
little. This reflects the
close similarity between the valence dielectric response (i.e.,
$\epsilon_2$ and $-{\rm Im}\, \epsilon^{-1}$) of water and ice,
and leads to an overall stretch of the spectra by about 1~eV. 
Unlike full GW approximations however, the MPSE fails to fully correct the 
gap in these systems, and the optical spectra that follow have been
aligned to match the observed gap.

\section {Ice Spectra}

Water has a rich phase diagram with at least 12 crystalline phases of ice, 9 of which are stable, 
in addition to several amorphous phases. Complicating the discussion of
H$_2$O, ice is proton disordered, but this cannot be treated exactly
by our approach which enforces periodic boundary conditions. 
Of the known, stable phases of ice, only ice-VIII has proton 
ordering with an anti-ferroelectric unit cell. Recent investigations
of the proton ordering phase transition between ices VII and VIII 
found no significant difference in the Oxygen K-edge NRIXS, implying that
the effect of proton disorder on that spectrum is small.\cite{Pylkkaenen2010}
%%%%%%%%%not sure why NIST reviewer circled this

In addition to local fluctuations in the dipole moments of most phases of ice, 
the low-mass hydrogen atoms have a significant amount of zero-point 
motion on top of thermal vibrations, which is not captured by standard Born-Oppenheimer 
molecular dynamics. Once again, accurate treatment of this disorder requires large 
unit cells which are not feasible in our approach. Recent work has focused on 
capturing the quantum nature of the hydrogen movement, but such improvements 
are not included here.\cite{PhysRevLett.101.017801}

\subsection{Ice Models}

For this study we have focused on two 
forms of ice; ice-VIII, which is stable below 273~K and pressures in the range 
of 2~GPa to 50~GPa, and ice-Ih, which is the common form of ice, stable under ambient pressure.
Our ice-Ih model uses a 16-molecule cell determined by enforcing the
experimental density and following the ice rules for
hydrogen placement with an O-H bond distance and angle of
%%%%%%% NIST reviewer didn't like Angstrom?
1.01~\AA\, and 106$^\circ$ respectively. 
A realistic simulation of ice-Ih should have full proton
disorder, but our models are based on small cells with periodic
boundary conditions, and thus contain artificial order. Further
calculations for other ice-Ih models using larger unit cells 
and finite temperature configuration sampling are called
for to assess the validity of our simplified model.

Ice-VIII is formed under high pressure 
at temperatures below the conventional freezing point of water. 
The pressure 
requirements preclude experimental techniques incompatible with 
diamond-anvil cells such as lower energy XAS measurements. However,
NRIXS can probe the same final-state 
as XAS with an additional tunable parameter $\mathbf{q}$,
the momentum transfer and at a much higher beam energy. The structure 
of ice-VIII is that of two interpenetrating cubic ice lattices resulting in 
a compressed second shell and higher densities than ice-Ih.
The unit cell contains only 4 molecules and has anti-ferroelectric Hydrogen ordering, 
making it an easier system to simulate than standard ice. Our ice-VIII cell uses the experimental 
lattice constants at a pressure of 2.4~GPa.\cite{kuhs1984}

\begin{figure}
\includegraphics[scale=0.3,angle=270]{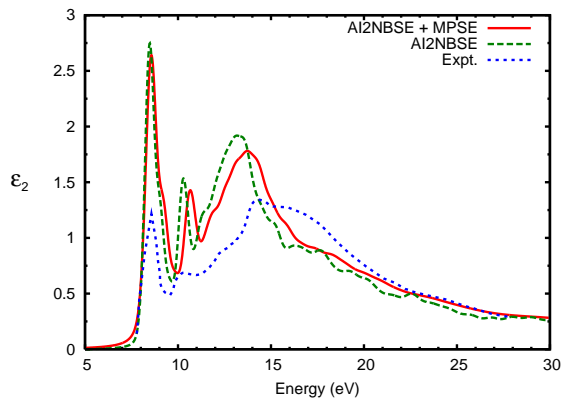}
\caption{(Color online)  The imaginary part of the dielectric constant
of ice-Ih calculated using AI2NBSE with the MPSE correction (solid red), 
and with a static energy shift (dashed green). For comparison an
experimental measurement of ice-Ih is shown (dotted blue).\cite{JPSJ.50.2643}
The discrepancy likely reflects a lack of disorder in our model structure
(see text).
}
\label{ice_eps2}
\end{figure}

\subsection{Ice valence spectra}

For the ice-Ih spectra,
the wavefunctions from ABINIT
were calculated on a $4\times4\times4$ $k$-point mesh with 464 bands
(400 conduction bands). 
For comparison, the structure and valence spectrum of hexagonal ice (ice-Ih)
were studied previously with DFT/BSE approaches, yielding qualitative 
agreement with experiment.\cite{PhysRevLett.94.037404}
For the valence spectrum (Fig.\ \ref{ice_eps2}), the calculated excitonic
peaks at 8~eV and 10~eV are significantly stronger than those measured.
Between 10~eV and 20~eV the absorption is also too strong and too thinly
peaked, compared to experiment which exhibits additional broadening past 15~eV.
These discrepancies are partly due to the lack of disorder in our
model structure, and point to the need to consider more elaborate models
that can account for such disorder. The effect of the MPSE 
is seen as a stretch of the spectra, together with energy dependent broadening.
Coupling between electronic excitations and phonons is expected to
provide small additional broadening, but is not included in this work.

\subsection{Ice core spectra}

\begin{figure}
\includegraphics[scale=0.3,angle=270]{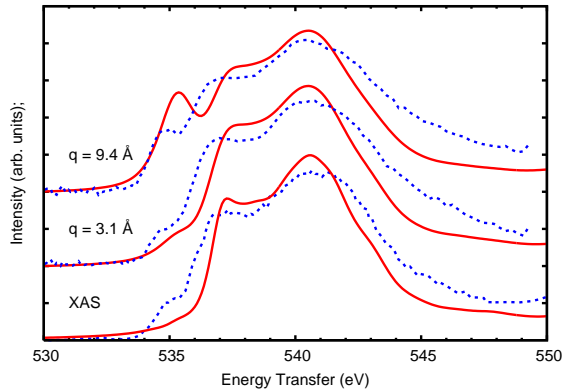}
\caption{ The calculated momentum-transfer dependence of the oxygen
K-edge in ice-Ih (solid red), including the MPSE. For comparison experimental results 
(dashed blue) for
XAS \cite{tse2008} and NRIXS
 \cite{Pylkkaenen2010} are shown.}
\label{Ih_xray}
\end{figure}

For the oxygen K-edge XAS and NRIXS calculations
a $4\times4\times4$ $k$-point mesh and 400 conduction bands were used 
for the BSE calculation and 900 bands for the screening. 
Oxygen K-edge XAS and NRIXS results for ice-Ih are shown in Fig.\
\ref{Ih_xray}. All calculated x-ray spectra have been broadened 
by convoluting with a Lorentzian to account for 
the core-hole lifetime and a Gaussian to match reported experimental broadening.
Further damping is provided by the imaginary part of the self-energy
(cf.\ Fig.\ \ref{SE_comp}). 
Results are shown averaged over orthogonal 
incident photon polarizations or momentum transfers for XAS and NRIXS respectively, 
since experimental data is reported for polycrystalline samples.

Our XAS calculation matches experiment fairly well, but has a lower
pre-edge (535~eV) 
and is noticeably too narrow in overall width (Fig.\ \ref{Ih_xray}).
Recently the momentum dependence of the 
O K-edge in ice has been measured using NRIXS.\cite{Pylkkaenen2010}
The theoretical simulation of NRIXS with
OCEAN differs only slightly from XAS in that the finite momentum
transfer ${\bf q}$ can break the dipole selection rules, allowing
transitions to s-type and higher angular momentum final states. 
Overall the calculations yield fair agreement for
the spectra, including both the relative weights
and the momentum dependence of the pre-edge feature 
(Fig.\ \ref{Ih_xray}). However, the balance between the s- and
p-character of the 
pre-edge is shifted too much towards the s-type in our calculation,
leading to overly strong growth of the pre-edge with increasing ${q}$.
Note that experiment shows little difference between ${q}\approx0$
(XAS) and ${q}=3.1$.

\begin{figure}
\includegraphics[scale=0.3,angle=270]{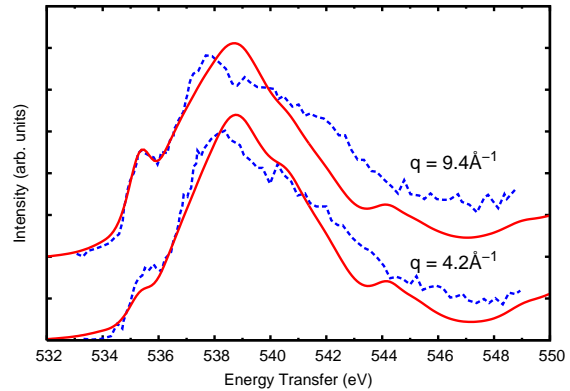}
\caption{(Color online)  The calculated (solid red) momentum transfer
dependence of the oxygen K-edge
in ice-VIII NRIXS
compared with experiment (dashed blue)\cite{Pylkkaenen2010} for 
small (${q}=4.2$~\AA$^{-1}$) and
large (${q}=9.4$~\AA$^{-1}$) momentum transfer.}
\label{8_nrixs}
\end{figure}

We have also examined 
the momentum dependence of ice-VIII (Fig.\ \ref{8_nrixs}) which exhibits
a similar 
evolution with increasing ${q}$ as ice-Ih, and good general agreement
with experiment. Like ice-Ih, our calculations of
the O K-edge in ice-VIII are too strongly peaked in the main edge
and have too little weight above about 543~eV when compared
to experiment, but show improvements over previous theoretical
results.\cite{Shaw2007} Similar to ice-Ih, the calculations of ice-VIII
also show excessive $\mathbf{q}$ dependence 
for the pre-edge, thus suggesting either a limitation of the DFT wave
functions or the need to consider thermal and quantum disorder.

\section{ Water: 17-molecule liquid cells}

\subsection{Water structural model}

Our calculations for optical and x-ray spectra were carried out on 
17-molecule snapshots for liquid water
obtained from the molecular dynamics (MD) results of Garbuio et
al.\cite{PhysRevLett.97.137402} 
Despite the small size of these model structures, they already exhibit
substantial disorder and are found to have a 
pair-distribution function (Fig.\ \ref{DoS})
in fairly good agreement with experiment. However, the O-O 
nearest neighbor distance is notably too long by a few tenths
of an angstrom.  Additionally, while bulk water is a disordered system 
with no net dipole moment, the limited cell size of our systems precludes 
this constraint; for our samples
the moments were found to vary widely from 14~Debye to 28~Debye. 

\begin{figure}
\includegraphics[scale=0.3,angle=270]{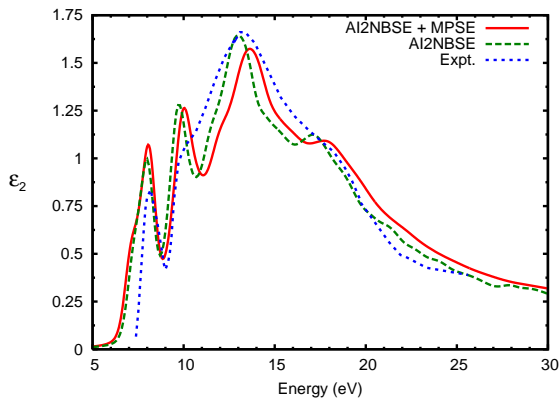}
\caption{(Color online) 
Calculated $\epsilon_2$ spectra for a single MD snapshot
with a 17-molecule water cell using
the MPSE (solid red) and a static energy shift (dashed green) compared to
experiment (dotted blue).\cite{Heller1974} }
\label{17_val}
\end{figure}

\subsection{Water valence spectra} 

For the valence calculation a $4\times4\times4$ $k$-point mesh and 300
conduction bands were used, and both a static energy shift
of 4.13 eV, as suggested by Garbuio et al.,\cite{PhysRevLett.97.137402} 
as well as the energy dependent MPSE were applied.
With the static shift the positions of the initial peaks match well
with experiment (Fig.\ \ref{17_val}) and in general, the agreement is excellent
over the full range of the spectra. Our results
for the peak positions in $\epsilon_2$ agree qualitatively with those of
Garbuio et al., but yield features that are smaller in overall magnitude.
The energy dependence of our MPSE (Fig.\ \ref{SE_comp})
introduces a stretch in the spectrum of $\approx$~5~\% in the range of 10~eV
to 20~eV, while Garbuio et al.\ reported the GW energy correction to be
almost constant across the low-lying conduction bands for water.

\subsection{Water core spectra}
%%%%%%%%%% Should we remove the error bars?
%%%%%%%%%% Add in comparison of recent K-edge cals?
\begin{figure}
\includegraphics[scale=0.3,angle=270]{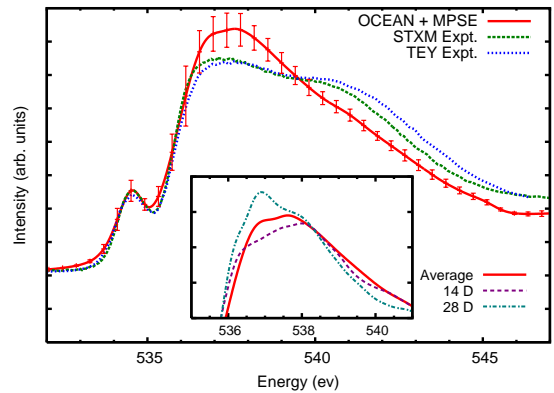}
\caption{(Color online) The XANES spectra for the O K-edge in water.
The OCEAN results, including the many-pole self-energy,
 are compared with recent Scanning Transmission X-ray Microscopy
(STXM)\cite{nilsson-2010} and Total Electron Yield (TEY)
experiments.\cite{Cappa2008} 
 The error-bars for the calculation represent the rms variation between
the 8 different MD snapshots, showing reasonable similarity between
different geometries sampled by the MD despite the limited cell size.
 Inset: The main edge for the MD snapshots with the lowest and highest
dipole moments compared to average (solid). 
 With decreasing dipole moment the first peak intensity
decreases with a slight shift to higher energies,
a trend consistent with
a decrease in the discrepancy with experiment as the net dipole moment 
approaches zero.
} 
\label{17_OK}
\end{figure}

XAS calculations were carried out for 8 of the MD snapshots using a 
$4\times4\times4$ $k$-point mesh including 300 bands 
for the final states and 400 conduction bands for the screening. The screened core-hole potential 
was found to be nearly identical for different O sites within a cell,
reflecting the molecular character
of liquid water.
The O K-shell XAS calculation (Fig.\ \ref{17_OK}) exhibits considerable
variation among the different oxygen sites both within a cell and between cells. 
Both the calculated and recent
experimental spectra\cite{nilsson-2010,Cappa2008}
exhibit a notable intensity shift and increased broadening going from ice
to water, shifting spectral weight to the lower part of the spectrum from 534~eV to 538~eV.
The calculated O K-edge spectra is in good agreement with experiments 
when corrected with the MPSE, but, as for the calculated ice-Ih spectra, 
the main edge at about 536~eV is stronger than 
in experiment. 
The peak at 541~eV in experiment is evident in the calculation but is
lacking in strength.
The success of the MPSE in correcting the overall width of the near-edge
is consistent with calculations using the COHSEX
approximation,\cite{chen-2009} but in contrast to earlier work
\cite{PhysRevLett.97.137402} which showed that 
the self-energy for water was nearly constant for their
$G^0W^0$ approximation,

In a disordered system like liquid water, each oxygen is subject 
to a different static potential and to slightly different core-hole screening 
by the valence electrons, resulting in shifts in the binding energy of the 
1s electrons. Exact, absolute energies were not calculated, but relative shifts 
were determined according to the relation
\begin{equation}
E_{1s} = E^{av}_{1s} + V^{KS} + \frac{1}{2} W_C. %( \Delta W_C - \Delta V_C  ) .
\end{equation}
$E^{av}_{1s}$ is the site-independent binding energy which has been aligned
to match experiment. The second term $V^{KS}$, is the total Kohn-Sham
potential at the site, including the effects of all the other ionic cores.
The last term is the effect of the spectator electrons screening the core
hole, reducing the energy necessary to excite the 1s electron. Both $V^{KS}$
and $W_C$ are evaluated at the Oxygen site. These binding-energy shifts
lead to a slight broadening of the spectra in Fig.\ \ref{17_OK}

We also find that the unphysical dipole moments of
the small unit-cells in this study have a systematic effect
on the calculated spectra (inset of Fig.\ \ref{17_OK}). That is, the
two main peaks of the calculated O K-edge at 537~eV and 540~eV exhibit
shifts in intensity and position
that loosely correlate with dipole moment, i.e.,.\ the local, static electric field felt by each water monomer 
in the ground state. At higher values of the dipole moment 
the main edge features shift to lower energies, increasing discrepancy with
experiment.  This again points to the need for
simulations with larger cells that more accurately characterize
the model structure and properties.

\section{Summary and Future Prospective}

\noindent

Theoretical calculations have been carried out for both valence and core
x-ray spectra of a number of model ice and water systems using modern
GW/BSE theoretical methods.  Our results suggest that accurate calculations
of the spectra for these systems are quite sensitive both to the theoretical
methods and details of the model structures. Nevertheless, we have shown
that the improved 
treatment of many-body effects in the present approach yields better agreement 
with experiment in terms of both relative peak weights as well as overall
width and feature locations compared to calculations that ignore these effects.
% Nevertheless, the improved treatments of many-body effects in
%the present approach generally yield improved agreement with experiment
%compared to calculations that ignore these effects.

 We find that the inclusion of an accurate quasi-particle self-energy
is important to characterize the damping and self-energy shifts in the
spectra. In particular we find that the stretch provided by our GW MPSE 
significantly improves the agreement between 
calculations and experiment for the ice and water systems.
Our results for the valence $\epsilon_2$ spectra for the 17-molecule water
model differ somewhat from those of Garbuio et al. Theirs have somewhat
larger values of $\epsilon_2$, while ours exhibits an extra excitonic peak.
The origin of these differences is likely due to differences in
the screening approximations used. In any case, our present
results appear to be in reasonable agreement with recent experimental results
both for the core- and valence-spectra of water.   However,
we suggest that configurational averages should be carried out for
ice-Ih and ice-VIII to better understand the effects of disorder,
finite temperature, and zero-point variations in structure.
A greater variation in hydrogen positioning and bonding 
from finite temperature and zero-point motion effects could lead to changes 
in feature weights as well as adding an increased broadening from disorder.
Additionally, convergence should be checked 
with respect to the size of the cells being used, especially because larger
cells should allow for better control of the net dipole moment. At present,
however, system size is limited by our codes and hence significantly
larger cells are computationally impractical.
%such calculations are impractical with our codes.  
The oxygen K-edge should be calculated
for a sufficient number of MD snapshots to ensure a good representation of the physical system.
Site-specific geometrical information, i.e., bond length,
angles, and hydrogen bond coordination, can then be compared to 
contributions from individual oxygen XANES
to correlate differences in local environment and calculated XAS/NRIXS.

ACKNOWLEDGMENTS: We thank R. Car, G. Galli, D. Prendergast, F. Gygi,
O. Pulci, L. Pettersson, A. Nilsson, and M. Ljungberg for
many helpful and stimulating discussions.
We are especially grateful to O. Pulci for supplying the
17-molecule water cells used in this work and to L. Pettersson and A. Nilsson
for making available recent experimental data. This work was supported in
part by DOE Grant DE-FG03-97ER45623. One of us (JJR) also wishes to thank
to acknowledge the Kavli Institute for Theoretical Physics Santa-Barbara
and the Ecole Polytechnique (Palaiseau, France)  for hospitality during
the completion of this work.

%\bibliographystyle{apsrev}
%\bibliography{ice.bib}

\end{document}